\documentclass[12pt] {article}
\begin{document}
\setcounter{page}{1}
\def\theequation{\arabic{section}.\arabic{equation}}
\def\theequation{\thesection.\arabic{equation}}
\setcounter{section}{0}

\title{On the D--wave state component of the deuteron in the
Nambu--Jona--Lasinio model of light nuclei}

\author{A. N. Ivanov~\thanks{E--mail: ivanov@kph.tuwien.ac.at, Tel.:
+43--1--58801--14261, Fax: +43--1--58801--14299}~${\textstyle
^\ddagger}$ , V. A. Ivanova~${\textstyle ^\ddagger}$,
H. Oberhummer~\thanks{E--mail: ohu@kph.tuwien.ac.at, Tel.:
+43--1--58801--14251, Fax: +43--1--58801--14299} ,\\
N. I. Troitskaya~\thanks{Permanent Address: State Technical
University, Department of Nuclear Physics, 195251 St. Petersburg,
Russian Federation} , M. Faber~\thanks{E--mail:
faber@kph.tuwien.ac.at, Tel.: +43--1--58801--14261, Fax:
+43--1--58801--14299}~\thanks{The authors are ordered according to
Russian alphabet as usual for publications [1,2,3]}}

\date{\today}

\maketitle

\begin{center}
{\it Atominstitut der \"Osterreichischen Universit\"aten,
Arbeitsbereich Kernphysik und Nukleare Astrophysik, Technische
Universit\"at Wien, \\ Wiedner Hauptstr. 8-10, A-1040 Wien,
\"Osterreich }
\end{center}

\begin{center}
\begin{abstract}
The D--wave state component of the neutron--proton bound state in the
deuteron is calculated in the Nambu--Jona--Lasinio model of light
nuclei.  For the ratio of D-- to S--state deuteron wave functions we
obtain equal to $\eta_{\rm d} = 0.0238$. This agrees well with the
phenomenological value $\eta_{\rm d} = 0.0256 \pm 0.0004$ quoted by
Kamionkowski and Bahcall (ApJ. {\bf 420}, 884 (1994)).
\end{abstract}
\end{center}

\begin{center}

\end{center}

\newpage

\section{Introduction}
\setcounter{equation}{0}

\hspace{0.2in} The Nambu--Jona--Lasinio model of light nuclei or
differently the nuclear Nambu--Jona--Lasinio (NNJL) model suggested in
[1--3] represents a relativistically covariant quantum field theoretic
approach to the description of low--energy properties and interactions
of the deuteron and light nuclei. The NNJL model is fully motivated by
QCD [1]. The deuteron appears {\it in the nuclear phase of QCD} as a
neutron--proton collective excitation, the Cooper np--pair, induced by
a phenomenological local four--nucleon interaction. The NNJL model
describes low--energy nuclear forces in terms of one--nucleon loop
exchanges providing a minimal transfer of nucleon flavours from
initial to final nuclear states and accounting for contributions of
nucleon--loop anomalies which are completely determined by
one--nucleon loop diagrams. The dominance of contributions of
nucleon--loop anomalies to effective Lagrangians of low--energy
nuclear interactions is justified in the large $N_C$ expansion, where
$N_C$ is the number of quark colours.

Nowadays there is a consensus concerning the existence of
non--nucleonic degrees of freedom in nuclei [4]. The non--nucleonic
degrees of freedom can be described either within QCD in terms of
quarks and gluons [5] or in terms of mesons and nucleon resonances
[6]. In the NNJL model the non--nucleonic degrees of freedom of nuclei
have been investigated in terms of the $\Delta(1232)$ resonance and
calculated the contribution of the $\Delta\Delta$ component to the
deuteron [2]. The obtained result $P(\Delta\Delta) = 0.3\,\%$ agrees
well with the experimental upper bound $P(\Delta\Delta) < 0.4\,\%$ [7]
and other theoretical estimates [4].

As has been shown in [3] the NNJL model describes well low--energy
nuclear forces for electromagnetic and weak nuclear reactions with the
deuteron of astrophysical interest such as the neutron--proton
radiative capture n + p $\to$ D + $\gamma$, the solar proton burning p
+ p $\to$ D + e$^+$ + $\nu_{\rm e}$, the pep--process p + e$^-$ + p
$\to$ D + $\nu_{\rm e}$ and reactions of the disintegration of the
deuteron by neutrinos and anti--neutrinos caused by charged $\nu_{\rm
e}$ + D $\to$ e$^-$ + p + p, $\bar{\nu}_{\rm e}$ + D $\to$ e$^+$ + n +
n and neutral $\nu_{\rm e}(\bar{\nu}_{\rm e})$ + D $\to$ $\nu_{\rm
e}(\bar{\nu}_{\rm e})$ + n + p weak currents.

The important problem which has not been jet clarified in the NNJL
model is related to the value of the contribution of the $D$--wave
state to the wave function of the deuteron. In this paper we fill this
blank. In section 2 we calculate the contribution of the D--wave state
to the wave function of the deuteron. We use a relativistically
covariant partial--wave analysis developed by Anisovich {\it et al.}
[8] for the description of nucleon--nucleon scattering.  The fraction
of the D--wave state of the deuteron wave function relative to the
S--wave one we obtain equal to $\eta_d = 0.0238$. This agrees well
with the value $\eta_d = 0.0256 \pm 0.0004$ quoted by Kamionkowski and
Bahcall [9]\,\footnote{The value $\eta_d = 0.0256 \pm 0.0004$ was
taken by Kamionkowski and Bahcall from Ref. [10].} who used this
parameter for the phenomenological description of the realistic wave
function of the deuteron in connection with the calculation of the
astrophysical factor $S_{\rm pp}(0)$ for the solar proton burning p +
p $\to$ D + e$^+$ + $\nu_{\rm e}$ in the potential model approach. In
the Conclusion we discuss the obtained result.

\section{The D--wave state component of the deuteron}
\setcounter{equation}{0}

\hspace{0.2in} The calculation of the value of the D--wave state
contribution to the wave function of the deuteron we would carry out
in terms of the amplitude of the transition n + p $\to$ D. We show
that the neutron--proton pair couples to the deuteron in both the
S--wave state and the D--wave state with the fraction of the D--wave
state agreeing with low--energy nuclear phenomenology.

In the NNJL model the phenomenological Lagrangian of the ${\rm n p D}$
interaction is defined by [1]
\begin{eqnarray}\label{label2.1}
{\cal L}_{\rm npD}(x) &=& - ig_{\rm V}[\bar{p^c}(x)\gamma^{\mu}n(x) -
\bar{n^c}(x)\gamma^{\mu}p(x)] D^{\dagger}_{\mu}(x)\nonumber\\ && +
\frac{g_{\rm T}}{2M_{\rm N}}[\bar{p^c}(x)\sigma^{\mu\nu}n(x) -
\bar{n^c}(x)\sigma^{\mu\nu}p(x)] D^{\dagger}_{\mu\nu}(x) + {\rm h.c.}
\end{eqnarray}
where $D^{\dagger}_{\mu}(x)$, $n(x)$ and $p(x)$ are the interpolating
fields of the deuteron, the neutron and the proton,
$D^{\dagger}_{\mu\nu}(x) = \partial_{\mu}D^{\dagger}_{\nu}(x) -
\partial_{\nu}D^{\dagger}_{\mu}(x)$ is the deuteron field strength.
The phenomenological coupling constant $g_{\rm V}$ is related to the
electric quadrupole moment of the deuteron $Q_{\rm D} = 0.286\,{\rm
fm}^2$, $g^2_{\rm V} = 2\pi^2 Q_{\rm D}M^2_{\rm N}$ [1], where $M_{\rm
N} = 940\,{\rm MeV}$ is the nucleon mass. The coupling constants
$g_{\rm V}$ and $g_{\rm T}$ are connected by the relation [1]
\begin{eqnarray}\label{label2.2}
g_{\rm T} = \sqrt{\frac{3}{8}}\,g_{\rm V},
\end{eqnarray}
which is valid at leading order in the large $N_C$ expansion [1].

The amplitude of the transition n + p $\to$ D is determined by
\begin{eqnarray}\label{label2.3}
\langle k_{\rm D},\lambda_{\rm D}|{\cal L}_{\rm npD}(0)|k_{\rm
p},\sigma_{\rm p}; k_{\rm n},\sigma_{\rm n}\rangle = \frac{M({\rm
n(k_{\rm n},\sigma_{\rm n}) + p(k_{\rm p},\sigma_{\rm p}) \to
D}(k_{\rm D},\lambda_{\rm D}))}{\sqrt{2E_{\rm D}V2E_{\rm n}V2E_{\rm
p}V}} ,
\end{eqnarray}
where $(E_{\rm D}, k_{\rm D},\lambda_{\rm D})$, $(E_{\rm p}, k_{\rm
p},\sigma_{\rm p})$ and $(E_{\rm n}, k_{\rm n},\sigma_{\rm n})$ are
energies, 4--momenta and polarizations of the deuteron, the proton and
the neutron, respectively, $V$ is a normalization space volume. The
wave functions of the initial and the final states of the transition n
+ p $\to$ D are given by
\begin{eqnarray}\label{label2.4}
|k_{\rm p},\sigma_{\rm p}; k_{\rm n},\sigma_{\rm n}\rangle &=&
a^{\dagger}_{\rm p}(k_{\rm p},\sigma_{\rm p})\,a^{\dagger}_{\rm
n}(k_{\rm n},\sigma_{\rm n})|0\rangle,\nonumber\\ \langle k_{\rm
D},\lambda_{\rm D}| &=& \langle 0|a_{\rm D}(k_{\rm D},\lambda_{\rm
D}),
\end{eqnarray}
where $a^{\dagger}_{\rm p}(k_{\rm p},\sigma_{\rm p})$ and
$a^{\dagger}_{\rm n}(k_{\rm n},\sigma_{\rm n})$ are creation operators
of the proton and the neutron, $a_{\rm D}(k_{\rm D},\lambda_{\rm D})$
is the annihilation operator of the deuteron and $|0\rangle$ is a
vacuum wave function. The relativistically invariant amplitude $M({\rm
n(k_{\rm n},\sigma_{\rm n}) + p(k_{\rm p},\sigma_{\rm p}) \to
D}(k_{\rm D},\lambda_{\rm D}))$ reads
\begin{eqnarray}\label{label2.5}
\hspace{-0.3in}&&M({\rm n(k_{\rm n},\sigma_{\rm n}) + p(k_{\rm
p},\sigma_{\rm p}) \to D}(k_{\rm D},\lambda_{\rm D})) =
e^{*\,\nu}(k_{\rm D},\lambda_{\rm D})\nonumber\\
\hspace{-0.3in}&&\times\,\Bigg\{2ig_{\rm V}[\bar{u^c}(k_{\rm
n},\sigma_{\rm n})\gamma_{\nu}u(k_{\rm p}, \sigma_{\rm p})] -
\frac{2ig_{\rm T}}{M_{\rm N}}[\bar{u^c}(k_{\rm n},\sigma_{\rm
n})\sigma_{\mu\nu}u(k_{\rm p},\sigma_{\rm p})]\,(k_{\rm n} + k_{\rm
p})^{\mu}\Bigg\},
\end{eqnarray}
where $\bar{u^c}(k_{\rm n},\sigma_{\rm n})$ and $u(k_{\rm p},
\sigma_{\rm p})$ are bispinorial wave functions of the neutron and the
proton with 4-momenta $k_{\rm n}$, $k_{\rm p}$ and polarizations
$\sigma_{\rm n}$, $\sigma_{\rm p}$; $e^{*\,\nu}(k_{\rm D},\lambda_{\rm
D})$ is a 4--vector of polarization of the deuteron with a 4--momentum
$k_{\rm D}$ and polarization $\lambda_{\rm D}$. The 4--momenta $k_{\rm
D}$, $k_{\rm n}$ and $ k_{\rm p}$ are related by $k_{\rm D} = k_{\rm
n} + k_{\rm p}$ due to conservation of energy and momentum.

As has been shown by Anisovich {\it et. al.} [8] for neutron--proton
scattering the neutron--proton densities describing the S-- and
D--wave states of a neutron--proton pair are equal to
\begin{eqnarray}\label{label2.6}
\Psi_{\nu}({^3}{\rm S}_1; \sigma_{\rm n}, \sigma_{\rm p})&=&
[\bar{u^c}(k_{\rm n},\sigma_{\rm n}){\cal S}_{\nu}u(k_{\rm p},
\sigma_{\rm p})],\nonumber\\ \Psi_{\nu}({^3}{\rm D}_1; \sigma_{\rm n},
\sigma_{\rm p})&=& [\bar{u^c}(k_{\rm n},\sigma_{\rm n}){\cal
D}_{\nu}u(k_{\rm p}, \sigma_{\rm p})],
\end{eqnarray}
where ${\cal S}_{\nu}$ and ${\cal D}_{\nu}$ are relativistically
covariant operators of the projection onto the S--wave  and the
D--wave state, respectively [8]:
\begin{eqnarray}\label{label2.7}
{\cal S}_{\nu} &=&\frac{1}{\sqrt{2s}}\,\Bigg[\gamma^{\perp}_{\nu} -
\frac{2k_{\nu}}{2M_{\rm N} + \sqrt{s}}\Bigg],\nonumber\\ {\cal
D}_{\nu} &=&\frac{2}{s^{3/2}}\,\Bigg[\frac{1}{4}\,(4M^2_{\rm N} -
s)\gamma^{\perp}_{\nu} - (M_{\rm N} + \sqrt{s})\,k_{\nu}\Bigg].
\end{eqnarray}
Here $P = k_{\rm p} + k_{\rm n}$, $k= \frac{1}{2}\,(k_{\rm p} - k_{\rm
n})$, $s = P^2$, $P\cdot k = 0$ and
\begin{eqnarray}\label{label2.8}
\gamma^{\perp}_{\nu} = \gamma_{\nu} - \hat{P}\,\frac{P_{\nu}}{s}.
\end{eqnarray}
The neutron--proton densities Eq.(\ref{label2.6}) are normalized by
the condition [8]
\begin{eqnarray}\label{label2.9}
\hspace{-0.3in}&&\frac{1}{3}\int {\rm tr}\{L_{\mu}(\hat{k}_{\rm p} + M_{\rm
N})L^{\mu}(-\hat{k}_{\rm n} + M_{\rm N})\}\,(2\pi)^4\delta^{(4)}(P -
k_{\rm p} - k_{\rm n})\,\frac{d^3k_{\rm p}}{(2\pi)^3 2E_{\rm p}}\,
\frac{d^3k_{\rm n}}{(2\pi)^3 2E_{\rm n}} = \rho_L(s),\nonumber\\
\hspace{-0.3in}&&
\end{eqnarray}
where $L_{\mu} = {\cal S}_{\mu}$ or ${\cal D}_{\mu}$, the factor 3 in
the denominator of the l.h.s.  describes the number of the states of a
neutron--proton density with a total momentum $J = 1$, $2J + 1 = 3$,
and $\rho_{\cal S}(s)$ and $\rho_{\cal D}(s)$ amount to
\begin{eqnarray}\label{label2.10}
\rho_{\cal S}(s) &=& \frac{1}{8\pi}\, \Bigg(\frac{s - 4M^2_{\rm
N}}{s}\Bigg)^{1/2},\nonumber\\ \rho_{\cal D}(s) &=& \frac{1}{8\pi}\,
\Bigg(\frac{s - 4M^2_{\rm N}}{s}\Bigg)^{5/2}.
\end{eqnarray}
In the center of mass frame of the neutron--proton pair the densities
Eq.(\ref{label2.6}) are equal to
\begin{eqnarray}\label{label2.11}
\hspace{-0.5in}\Psi_{0}({^3}{\rm S}_1; \sigma_{\rm n}, \sigma_{\rm
p})&=& [\bar{u^c}(k_{\rm n},\sigma_{\rm n}){\cal S}_0u(k_{\rm p},
\sigma_{\rm p})] = 0,\nonumber\\ \hspace{-0.5in}\vec{\Psi}({^3}{\rm S}_1; \sigma_{\rm
n}, \sigma_{\rm p})&=& [\bar{u^c}(k_{\rm n},\sigma_{\rm n})\vec{{\cal
S}}u(k_{\rm p}, \sigma_{\rm p})] =
\frac{1}{\sqrt{2}}\,\varphi^{\dagger}_{\rm n}(\sigma_{\rm
n})\vec{\sigma}\varphi_{\rm p}(\sigma_{\rm p}),\nonumber\\
\hspace{-0.5in}\Psi_{0}({^3}{\rm D}_1; \sigma_{\rm n}, \sigma_{\rm
p})&=& [\bar{u^c}(k_{\rm n},\sigma_{\rm n}){\cal D}_0u(k_{\rm p},
\sigma_{\rm p})] = 0,\nonumber\\ \hspace{-0.5in}\vec{\Psi}({^3}{\rm
D}_1; \sigma_{\rm n}, \sigma_{\rm p})&=& [\bar{u^c}(k_{\rm
n},\sigma_{\rm n})\vec{{\cal D}}u(k_{\rm p}, \sigma_{\rm p})] = -
\frac{1}{2}\,\varphi^{\dagger}_{\rm n}(\sigma_{\rm
n})[3\,(\vec{\sigma}\cdot \vec{v})\,\vec{v} -
\vec{v}^{\,2}\,\vec{\sigma}]\, \varphi_{\rm p}(\sigma_{\rm p}),
\end{eqnarray}
where $\vec{v} = \vec{k}/\sqrt{\vec{k}^{\,2} + M^2_{\rm N}} = \sqrt{1
- 4M^2_{\rm N}/s}$ and $\vec{k}$ are a relative velocity and a
3--momentum of the neutron--proton pair, $\varphi_{\rm n}(\sigma_{\rm
n})$ and $\varphi_{\rm p}(\sigma_{\rm p})$ are spinorial wave
functions of the neutron and the proton, respectively. It is obvious
that the densities Eq.(\ref{label2.11}) describe the neutron--proton
pair in the S-- and D--wave states with a total spin $S = 1$ and a
total momentum $J = 1$.

The neutron--proton densities Eq.(\ref{label2.11}) are normalized by
\begin{eqnarray}\label{label2.12}
\frac{1}{3}\sum_{\sigma_{\rm n} = \pm 1/2}\sum_{\sigma_{\rm p} = \pm
1/2}\vec{\Psi}^{\dagger}({^3}{\rm S}_1; \sigma_{\rm n}, \sigma_{\rm
p})\cdot \vec{\Psi}({^3}{\rm S}_1; \sigma_{\rm n}, \sigma_{\rm p}) &=&
1,\nonumber\\ \frac{1}{3}\sum_{\sigma_{\rm n} = \pm
1/2}\sum_{\sigma_{\rm p} = \pm 1/2}\vec{\Psi}^{\dagger}({^3}{\rm D}_1;
\sigma_{\rm n}, \sigma_{\rm p})\cdot\vec{\Psi}({^3}{\rm D}_1;
\sigma_{\rm n}, \sigma_{\rm p}) &=& v^{\,4} = \Bigg(1 -
\frac{4M^2_{\rm N}}{s}\Bigg)^{\!\!2}.
\end{eqnarray}
The decomposition of the neutron--proton densities in the amplitude
Eq.(\ref{label2.5}) into the densities with a certain orbital momentum
we would carry out at leading order in the large $N_C$ expansion
[1--3]. This would allow to consider the neutron and the proton as
free particles obeying free equations of motion
\begin{eqnarray}\label{label2.13}
\bar{u^c}(k_{\rm n},\sigma_{\rm n})(\hat{k}_{\rm n} + M_{\rm N}) &=&
0,\nonumber\\ (\hat{k}_{\rm p} - M_{\rm N})\,u(k_{\rm p}, \sigma_{\rm
p}) &=& 0.
\end{eqnarray}
In order to express the neutron--proton densities in the amplitude
Eq.(\ref{label2.5}) in terms of the projection operators
Eq.(\ref{label2.7}), first, we have to exclude the term containing
$\sigma_{\mu\nu}$. This can be carried out by using Gordon's identity
\begin{eqnarray}\label{label2.14}
\hspace{-0.3in}&&[\bar{u^c}(k_{\rm n},\sigma_{\rm n})
\sigma_{\mu\nu}u(k_{\rm p},\sigma_{\rm p})]\,\frac{(k_{\rm n} + k_{\rm
p})^{\mu}}{2M_{\rm N}} = - [\bar{u^c}(k_{\rm n},\sigma_{\rm
n})\gamma_{\nu}u(k_{\rm p},\sigma_{\rm p})] + \frac{k_{\nu}}{M_{\rm
N}}\,[\bar{u^c}(k_{\rm n},\sigma_{\rm n})u(k_{\rm p},\sigma_{\rm
p})].\nonumber\\
\hspace{-0.3in}&&
\end{eqnarray}
Substituting Eq.(\ref{label2.14}) in Eq.(\ref{label2.5}) we get
\begin{eqnarray}\label{label2.15}
\hspace{-0.3in}&&M({\rm n(k_{\rm n},\sigma_{\rm n}) + p(k_{\rm
p},\sigma_{\rm p}) \to D}(k_{\rm D},\lambda_{\rm D})) = 2i(g_{\rm V} +
2g_{\rm T})\,e^{*\,\nu}(k_{\rm D},\lambda_{\rm D})\nonumber\\
\hspace{-0.3in}&&\times\,\Bigg\{[\bar{u^c}(k_{\rm n},\sigma_{\rm
n})\gamma_{\nu}u(k_{\rm p}, \sigma_{\rm p})] - \frac{2g_{\rm
T}}{g_{\rm V} + 2g_{\rm T}}\,\frac{k_{\nu}}{M_{\rm
N}}\,[\bar{u^c}(k_{\rm n},\sigma_{\rm n})u(k_{\rm p},\sigma_{\rm
p})]\Bigg\}.
\end{eqnarray}
In terms of ${\cal S}_{\nu}$ and ${\cal D}_{\nu}$ vectors
$\gamma^{\perp}_{\nu}$ and $k_{\nu}$ are determined by
\begin{eqnarray}\label{label2.16}
\gamma^{\perp}_{\nu} &=& \frac{2\sqrt{2}}{3}\,(M_{\rm N} +
\sqrt{s})\,{\cal S}_{\nu} - \frac{2}{3}\,\frac{s}{2M_{\rm N} +
\sqrt{s}}\,{\cal D}_{\nu},\nonumber\\ k_{\nu} &=&
\frac{1}{3\sqrt{2}}\,(4M^2_{\rm N} - s)\,{\cal S}_{\nu} -
\frac{1}{3}\,s\,{\cal D}_{\nu}.
\end{eqnarray}
Substituting Eq.(\ref{label2.16}) in Eq.(\ref{label2.15}) and taking
into account that $[\bar{u^c}(k_{\rm n},\sigma_{\rm n})
\hat{P}u(k_{\rm p},\sigma_{\rm p})] = 0$ we obtain
\begin{eqnarray}\label{label2.17}
\hspace{-0.5in}&&M({\rm n(k_{\rm n},\sigma_{\rm n}) + p(k_{\rm
p},\sigma_{\rm p}) \to D}(k_{\rm D},\lambda_{\rm D})) =
4\sqrt{2}i(g_{\rm V} + 2g_{\rm T})\,M_{\rm N}\,e^{*\,\nu}(k_{\rm
D},\lambda_{\rm D})\nonumber\\
\hspace{-0.5in}&&\times\,\{[\bar{u^c}(k_{\rm n},\sigma_{\rm
n}){\cal S}_{\nu}u(k_{\rm p}, \sigma_{\rm p})] + \eta_{\rm
d}\,[\bar{u^c}(k_{\rm n},\sigma_{\rm n}){\cal D}_{\nu}u(k_{\rm
p},\sigma_{\rm p})]\} = \nonumber\\
\hspace{-0.5in}&&= 4\sqrt{2}i(g_{\rm V} + 2g_{\rm T})\,M_{\rm
N}\,e^{*\,\nu}(k_{\rm D},\lambda_{\rm D})\,[\Psi_{\nu}({^3}{\rm S}_1;
\sigma_{\rm n}, \sigma_{\rm p}) + \eta_{\rm d}\,\Psi_{\nu}({^3}{\rm
D}_1; \sigma_{\rm n}, \sigma_{\rm p})],
\end{eqnarray}
where $\eta_{\rm d}$ describes the fraction of the D--wave state in
the wave function of the deuteron. It is equal to
\begin{eqnarray}\label{label2.18}
\eta_{\rm d} = \frac{1}{3\sqrt{2}}\,\frac{ 2\,g_{\rm T} - g_{\rm
V}}{2\,g_{\rm T} + g_{\rm V}}.
\end{eqnarray}
For the derivation of Eqs.(\ref{label2.17}) and (\ref{label2.18}) we
have set $s = M^2_{\rm D}$ and neglected the contribution of the
binding energy of the deuteron in comparison with a nucleon mass
$M_{\rm N}$. This means that $4M^2_{\rm N} - s = 0$ when compared with
$M^2_{\rm N}$.

Using the relation Eq.(\ref{label2.2}) the parameter $\eta_{\rm d}$
takes the value
\begin{eqnarray}\label{label2.19}
\eta_{\rm d} = \frac{1}{3\sqrt{2}}\,\frac{\sqrt{3} -
\sqrt{2}}{\sqrt{3} + \sqrt{2}} = 0.0238.
\end{eqnarray}
This agrees well with the value $\eta_{\rm d} = 0.0256 \pm 0.0004$
that was used in low--energy nuclear phenomenology for the description
of the realistic wave function of the deuteron within the potential
model approach [9, 10].

\section{Conclusion}
\setcounter{equation}{0}

\hspace{0.2in} We have shown that the NNJL model describes well in
agreement with low--energy nuclear phenomenology [10] such a fine
structure of the deuteron as a contribution of the D--wave state. The
calculation of the fraction of the D--wave state to the wave function
of the deuteron we have carried out at leading order in the large
$N_C$ expansion [1]. This has allowed to treat the neutron and the
proton as free particles on--mass shell [2] obeying free equations of
motion.  To the decomposition of the amplitude of the transition n + p
$\to$ D into the neutron--proton quantum field configurations having
certain orbital momenta and corresponding to the S-- and D--wave
states, respectively, we have applied a relativistically covariant
partial--wave analysis invented by Anisovich {\it et al.} [8] for the
description of nucleon--nucleon scattering with nucleon--nucleon pairs
coupled in the states with certain orbital momenta.

The theoretical value of the D--wave state fraction in the wave
function of the deuteron $\eta_{\rm d} = 0.0238$ calculated in the
NNJL model agrees well with low--energy nuclear phenomenology giving
$\eta_{\rm d} = 0.0256 \pm 0.0004$ [10]. The former was quoted by
Kamionkowski and Bahcall [9] for the parameterization of the realistic
wave function of the deuteron in connection with the calculation of
the astrophysical factor $S_{\rm pp}(0)$ for the solar proton burning
p + p $\to$ D + e$^+$ + $\nu_{\rm e}$. The calculation of the
contribution of the D--wave state fraction of the wave function of the
deuteron in agreement with low--energy nuclear phenomenology testifies
that the NNJL model describes to full extent low--energy tensor
nuclear forces playing an important role in low--energy nuclear
physics on the whole and for the existence of the deuteron, in
particular [11].

\section*{Acknowledgement}

\hspace{0.2in} One of the authors (V. Ivanova) is grateful to her
supervisor Prof. A. A. Choban for discussions.

\end{document}